\documentclass[12pt]{iopart}
\begin{document}

\title[]{COMMENT}{Comment on ``On the importance of the free energy for 
elasticity under pressure''}

\author{Gerd Steinle-Neumann\dag
\footnote[3]{To whom correspondence should be addressed 
(g.steinle-neumann@uni-bayreuth.de)}, R. E. Cohen\ddag} 

\address{\dag\ Bayerisches Geoinstitut, Universit\"at Bayreuth, 95440 Bayreuth,
Germany}
\address{\ddag\ Geophysical Laboratory, Carnegie Institution of Washington, 
5251 Broad Branch Rd., NW, Washington, DC 20015, USA}

\begin{abstract}
Marcus et al. (Marcus P, Ma H and Qiu S L 2002 J. Phys.:
Condens. Matter 14 L525) claim that thermodynamic properties of materials
under pressure must be computed using the Gibbs free energy $G$, rather than 
the internal energy $E$. Marcus et al. state that ``The minima of $G$, but not of $E$, give 
the equilibrium structure; the second derivatives of $G$, 
but not of $E$, with respect to strains at the equilibrium structure give 
the equilibrium elastic constants.'' Both statements are incorrect.
\end{abstract}

\submitto{\JPCM}

\maketitle

Marcus et al.\ \cite{marcus} presented an analysis of
structural and elastic properties of solids subject to compression under
athermal ($T=0$) conditions. They claim that the Gibbs free energy $G$ must 
be used at finite pressure $P$ to find the equilibrium structure rather than 
the internal energy $E$. In particular, 
they consider the epitaxial Bain path (EBP) that relates body-centered cubic 
($bcc$), body-centered tetragonal ($bct$), and face-centered cubic ($fcc$) 
structures for Fe at 100 GPa (see also ref.\ \cite{mqm}). They show that 
while $G(c/a;p)$ along the EBP (with $a$ and $c$ the two independent lattice 
parameters) yields a minimum at $c/a=1$ (the bcc structure), the minimum of  
$E(c/a;p)$ along the EBP is displaced, at $c/a=0.95$. This is a result of 
misusing elementary thermodynamics \cite{pippard}. There is a minimum 
principle for the internal energy $E$ at constant entropy $S$ and volume $V$, 
for the enthalpy $H = E + P V$ at constant $S$ and $P$, for the Helmholtz 
free energy $A = E - T S$ at constant temperature $T$ and $V$, and for the 
Gibbs free energy $G = E - T S + P V$ at constant $P$ and $T$. There is no 
minimum principle 
for $E$ (or $H$, which are equivalent at $T = 0$, considered by Marcus et al.) 
at constant $P$. The correct analysis is shown in Fig.\ 3 of Stixrude et 
al. \cite{stixrude}, where $E$ versus $c/a$ is shown at constant $V$ for the 
EBP for Fe, and the extrema are at the $bcc$ structure, as expected. There is 
nothing wrong with minimizing $G$ at constant $P$ and $T$, but the exact same 
results will be obtained for minimizing $E$ at constant $V$ and $S$. The 
resulting pressure $P$ can always be obtained from $E$, since 
$P = -\frac{\partial E}{\partial V}\mid_S$. At $T=0$ the constraint $dS = 0$ 
is trivial, since $S=0$ at $T=0$. 

Marcus et al.\ further assert that proper elastic constants must be 
calculated from second Eulerian strain ($\varepsilon$) derivatives of $G$
(defined as $c_{ij}$ in \cite{marcus}) rather than the internal energy $E$ 
($\bar{c}_{ij}$). Calculating the shear elastic constants  
$c^{\prime}=(c_{11}-c_{12})/2$ and $c_{44}$ for $bcc$-iron using both 
$G$ and $E$ ($\bar{c}^{\prime}$ and $\bar{c}_{44}$) they find
a shear instability at 150 GPa using $G$ ($c^{\prime}$) but not 
for $E$ ($\bar{c}^{\prime}$), implying that previous computational estimates 
of elastic constants at pressure are incorrect, and that pressure corrections 
need to be applied. To the contrary, computation of elastic constants as 
prescribed by Marcus et al.\ gives incorrect results. Marcus et al. ignored 
the fact that the pressure and shear stresses vary as a function of strain. 
They did not obtain any thermodynamically valid second derivatives by their 
finite difference procedure, in which they computed 
$\frac{1}{V_0} \left(E_1(\varepsilon_{ij}) + P_0 V_1(\varepsilon_{ij}) 
- E_0 - P_0 V_0\right)$; they did not even obtain the derivatives 
$\frac{1}{V}\frac{\partial^2 G(P)}{\partial \epsilon_i \epsilon_j}
\mid_{\epsilon_{ k..l}}$, which are not in any case elastic constants, 
since the pressure and shear stresses vary with their deformation 
$\varepsilon_{ij}$. This problem remains even if the second order coefficients of a fit for a 
polynomial expansion in $\frac{1}{V_0} \left(E + P_0 V \right)$ is obtained.  The high 
pressure elastic constants computed using their procedures \cite{mqm,qm1,qm2,jm1,jm2} are 
incorrect, although the resulting errors may be small in some cases.

It is important to use the appropriate thermodynamic function for the 
appropriate conditions, and as any student of elementary thermodynamics knows, 
it is also important to keep track of what is being held constant for a given 
partial derivative. Elastic constants can be defined by various ways: 
(1) from the equations of motion (i.e.\ sound velocities), (2) as derivatives 
of stress with respect to strain, or (3) as second derivatives of the 
internal (giving the adiabatic elastic constants) or Helmholtz free energy 
(giving the isothermal elastic constants) with respect to strain, holding 
the other strains constant \cite{bk}. Different constants can also be derived 
depending on the use of finite or infinitesimal strain parameters. All of 
these definitions are equivalent at zero pressure, but differ under applied 
stress. The different definitions of elastic constants under applied stress 
remain sources of confusion \cite{gregoryanz_reply}. Under no conditions, 
however, is an elastic constant tensor properly defined as derivatives of $H$ 
or $G$ with respect to strains, holding the other strains constant. 

Kamb \cite{kamb} even comes to the conclusion
that {\it it is not possible usefully to associate a Gibbs free energy with 
a non-hydrostatically stressed solid}; this is similarly stated by Wallace 
\cite{wallace65}. The definition of elasticity on the basis of $G$ is 
not well-founded. A problem arises, for example, when considering phase equilibria of a fluid 
in contact with a crystal surface; the chemical potential of 
components in the fluid in equilibrium with the solid vary according to the 
crystal face for a stressed solid, indicating there is no unique definition 
of the Gibbs free energy for a stresses solid \cite{kamb}.

Marcus et al. state that the elastic constants in Refs. 
\cite{wasserman,soderlind,steinle} are incorrect and require pressure 
corrections. The elastic constants presented in Refs. 
\cite{wasserman,steinle,stixrude_sc,cohen} are the elastic constants for wave 
propagation, 
and these are most easily measured, and are important in seismology and 
other applications. 

For isotropic initial stress the elastic constants for acoustic wave propagation and 
stress-strain coefficients are equivalent (see section 5 in \cite{bk}). We will now 
illustrate that the expression of strain-energy density can give the same 
elastic constants as the stress-strain relations for volume conserving 
strains for a reference state with isotropic applied stress. We use the 
fourth rank tensor notation from \cite{bk} for the elastic constants 
$c_{ijkl}$ (the stress-strain coefficients). Consider the expression for 
strain-energy density from Barron and Klein \cite{bk}
\begin{equation}\label{emaster}
\frac{\Delta E}{V}=-p \varepsilon_{ii}+\frac{1}{2} \left( c_{ijkl}-\frac{1}{2}p
\left( 2\delta_{ij}\delta_{kl}-\delta_{il}\delta_{jk}-\delta_{jl}\delta_{ik} 
\right) \right)\varepsilon_{ij}\varepsilon_{kl},
\end{equation}
where $\delta_{ik}$ is the Kronecker delta. Evaluating this 
expression, for example, for $c_{1313}$ (corresponding to 
$\bar{c}_{55}=\bar{c}_{44}$ in Voigt notation, as used in ref. \cite{marcus}),
with the strain
\begin{equation}
\varepsilon(d)=\left(\begin{array}{ccc}
0 & 0 & d \\ 0 & 0 & 0 \\ d & 0 & 0 
\end{array}\right),
\end{equation}
with $d$ the strain amplitude, one does indeed obtain a pressure correction 
term
\begin{equation}
\frac{\Delta E}{V}2\left( c_{1313}+\frac{1}{2}p\right)d^2.
\end{equation}
This is the strain used in ref \cite{soderlind}, and their elastic constants 
should consequently be corrected for pressure to obtain wave propagation 
velocity.

However, by choosing specific volume conserving strains, such as those given 
in \cite{steinle,cohen}, corrections can be avoided. For the $c_{1313}$ 
example we apply the monoclinic strain
\begin{equation}
\varepsilon(d)=\left(\begin{array}{ccc}
0 & 0 & d \\
0 & d^2/\left( 1-d^2\right) & 0 \\ d & 0 & 0 
\end{array}\right),
\end{equation}
eq. \ref{emaster} becomes 
\begin{equation}
\frac{\Delta E}{V}=-\frac{d^2p}{1-d^2}
                  +2\left(c_{1313}+\frac{1}{2}p \right)d^2 
                  +\frac{1}{2}c_{2222}\frac{d^4}{(1-d^2)^2} 
                  +2c_{1322}\frac{d^3}{1-d^2}.
\end{equation}
For hexagonal and tetragonal systems $c_{1322}=0$ and the final term in the 
sum on the right hand side is zero. Expanding this into a series of $d$ yields
\begin{equation}
\frac{\Delta E}{V} = \delta^2 \left( 2 c_{1313} \right)+ O[\delta^4],
\end{equation}
without any pressure correction term. This is also true for the other strains
given in \cite{steinle,cohen}.

Single crystal elasticity is difficult to measure in high 
pressure experiments, especially for opaque materials such as metals which are
discussed in \cite{marcus}. However, advances in optical 
spectroscopy have made it possible to measure the Raman active phonon mode 
in hcp metals. This optical mode and the shear elastic constant $c_{1313}$ 
can be viewed as properties of the same, nearly continuous phonon branch in 
an extended Brillouin zone scheme, and can be related by a simple force 
constant model of phonon dispersion \cite{upadh}. Results obtained for 
Fe and Re \cite{steinle} compare very favorably with experimental 
estimates \cite{merkel,jephcoat} over a wide pressure range, 
corroborating that no pressure correction need be applied. Computed finite temperature elastic 
constants for Ta \cite{gulseren} also agree well with sound velocities obtained under shock 
conditions along the Hugoniot. 

To conclude, the internal energy $E$ in conjunction with its
volume $V$ and strain derivatives defines the thermodynamics and 
elasticity of a material completely, even under applied stress. 
Strain derivatives of the Gibbs free energy $G$, on the other hand, do not 
yield properly defined elastic constants.

We greatly appreciate helpful discussions with L. Stixrude.
The work was supported by the US National Science Foundation under grants
EAR-99080602 and EAR-0310139, and by the US Department of Energy ASCI/ASAP 
subcontract B341492 to 
Caltech DOE W-7405-ENG-48 to REC.

\end{document}